\documentclass{ws-procs9x6}

\begin{document}

\title{QUANTIFYING THE UNKNOWN: ISSUES IN SIMULATION VALIDATION AND THEIR
EXPERIMENTAL IMPACT}

\author{M. G. PIA $^*$, M. BATI\v{C}, G. HOFF and P. SARACCO}
\address{INFN Sezione di Genova,\\
Genova, 16136, Italy\\
$^*$E-mail: MariaGrazia.Pia@ge.infn.it\\
www.ge.infn.it}

\author{M. BEGALLI}
\address{State University of Rio de Janeiro,\\
Rio de Janeiro, RJ 20550-013, Brazil\\
E-mail: begalli@fnal.gov }

\author{M. HAN, C. H KIM and H. SEO }
\address{Hanyang University,\\
Seoul, 133-791, Korea\\
E-mail: chkim@hanyang.ac.kr }

\author{S. HAUF and M. KUSTER}
\address{Technical University Darmstadt, \\
Darmstadt, 64289, Germany,\\
E-mail: steffen.hauf@astropp.physik.tu-darmstadt.de }

\author{L. QUINTIERI}
\address{INFN Laboratori Nazionali di Frascati,\\
Frascati, 00044, Italy\\
E-mail: Lina.Quintieri@lnf.infn.it  }

\author{G. WEIDENSPOINTNER}
\address{Max-Planck-Institut Halbleiterlabor,\\
Munich, 81739, Germany\\
E-mail: Georg.Weidenspointner@hll.mpg.de  }

\author{A. ZOGLAUER}
\address{University of California at Berkeley,\\
Berkeley, CA 94720-7450, USA\\
E-mail: zog@ssl.berkeley.edu  }

\begin{abstract}

The assessment of the reliability of Monte Carlo simulations is discussed, with
emphasis on uncertainty quantification and the related impact on experimental
results. Methods and techniques to account for epistemic uncertainties, i.e. for
intrinsic knowledge gaps in physics modeling, are discussed with the support of
applications to concrete experimental scenarios. Ongoing projects regarding the
investigation of epistemic uncertainties in the Geant4 simulation toolkit are
reported.

\end{abstract}


\bodymatter

\section{Introduction}
\label{sec_intro}

The investigation and quantification of epistemic uncertainties \cite{epistemic}
is well established in the domain of deterministic simulation, but it is a
relatively new domain of research in the context of Monte Carlo simulation.
It concerns the issue of how epistemic uncertainties, i.e. uncertainties due to
lack of knowledge, namely in modeling physics processes, affect the outcome of
Monte Carlo simulation. 
Epistemic uncertainties are present in Monte Carlo codes, when the absence of
experimental data, or inconsistencies in available measurements, prevent the
achievement of firm conclusions regarding the correct values of physics
parameters or the validity of physics models used in the simulation.
Epistemic uncertainties can induce systematic effects in the simulation; this
issue is especially important, since  can negatively affect the accuracy and
reliability of simulation results.

Due to their intrinsic nature, related to lack of knowledge, epistemic
uncertainties are difficult to quantify.
Despite their importance in complex systems, there
is no generally accepted method of measuring epistemic uncertainties and they
contribute to the reliability of the whole system.
A variety of mathematical formalisms \cite{helton} has been developed for this
purpose; the most common methods adopted in the context of deterministic
simulations are interval analysis and applications of Dempster-Shafer theory of
evidence \cite{shafer}.
Nevertheless, these techniques may not always be directly applicable in
identical form to the treatment of epistemic uncertainties in Monte Carlo
simulations.

Sensitivity analysis \cite{what} is a tool for exploring how 
uncertainties influence the model output.
This approach is adopted in two exploratory projects, which intend to evaluate
possible methods for uncertainty quantification related to the Geant4
\cite{g4nim,g4tns} simulation toolkit.
Epistemic uncertainties are usually represented in statistical analyses as a set
of discrete possible or plausible choices; in the exploratory analyses described
here the possible choices concerned the values of physical parameters or a set
of alternative physics models.

\section{Proton depth dose simulation}

This study assesses the impact of epistemic uncertainties associated with
various physics models and parameters relevant to Monte Carlo codes through the
simulation of a concrete use case: the depth dose profile in water generated by
a proton beam as in a typical therapeutical facility.
For this purpose the geometry of a realistic hadrontherapy facility \cite{tns_pablo}
publicly available as a Geant4 example was utilized.

A sensitivity analysis has examined the response of the system to a wide set of
modeling approaches; this method plays a conceptually similar role to the
interval analysis method applied in deterministic simulation, where parameters
subject to epistemic uncertainties are varied within bounds.
The environment for this analysis has been realized in the context of a
Geant4-based application; the characteristic of Geant4 as a toolkit,
encompassing a wide variety of physics models, allow the configuration of the
simulation with a large number of different physics options in the same software
environment.
The outcome associated with the various models subject to investigation has been
compared by means of rigorous statistical analysis methods to quantitatively
estimate the effect of physics-related systematic uncertainties.

Epistemic uncertainties are associated with parameters used by the simulation
models: proton stopping powers and the water mean ionization potential, for
whose values a consensus has not yet been achieved in the scientific community.
The interval analysis has highlighted a shift in the
position of the Bragg peak related to range of variability of these parameters.

Nuclear interactions, both eleastic and inelastic, affect the shape of the depth
dose distribution.
Epistemic uncertainties in this domain derive from the still incomplete
validation of the hadronic models used by the simulation.
No statistically significant effects on the depth dose profile have been
identified as a result of the interval analysis; nevertheless, significant
systematic differences deriving from epistemic uncertainties are observed in
other features of the simulation outcome, such as secondary particle production.
Multiple scattering modeling also plays an important role in the evaluation of 
possible sources of systematic effects.

Sensitivity analysis as applied to this simulation topic contributes to identify
and quantify possible systematic effects in the simulation; it cannot infer
anything about the validity of any of the physics models, for which experimental
data would be needed.

The analysis of the proton depth dose simulation shows that the appearance of
systematic effects generated by epistemic uncertainties in the physic models
depends not only on the intrinsic characteristics of the uncertainties, but also
on the characteristics of the simulation environment.

\section{Atomic binding energies}

General purpose Monte Carlo codes use a variety of compilations of atomic
electron binding energies, either deriving from theoretical calculations or from
empirical evaluations of direct and indirect experimental data.
Despite the fundamental character of these atomic parameters, there is no
consensus among the various Monte Carlo systems and physics models about their
values: the differences across the binding energies reported in the various 
compilations range from the electronvolt scale to several hundred 
electronvonvolts.

The analysis adopted two complementary approaches: on one side direct validation
based on binding energies measurements, on the other side the evaluation of how
different compilations of these parameters contribute to the accuracy of physics
observables calculated by the simulation with respect to experimental data.

Reference experimental values for direct validation of atomic binding energies
are relatively limited: the main issue for direct validation consists of
discrepancies in experimental values due to calibration effects, for instance
when measurements are taken in different laboratories and exploit different
experimental techniques.
Two sets of reference data concerning core shells, which have been subject to a
process of recalibration and evaluation, have been assembled by Powell
\cite{powell} and NIST (United States National Institute of Standards)
\cite{nist_ref}, which encompass respectively only 65 and 81 binding energy values.
In addition, NIST reports reference ionization energies for all elements \cite{nist_ioni}.

Direct comparison of the binding energies in the various compilations by means
of statistical methods has identified the compilation by Williams \cite{williams} as the one,
among those considered in this study, exhibiting the best compatibility with
Powell and NIST reference data.
Regarding ionization energies, the compilation by Carlson \cite{carlson} appears to best
reproduce NIST reference values.
Characteristic X-ray energies are best reproduced by the compilation by Larkins \cite{larkins}.

The values of atomic binding energies can significantly affect the accuracy of
ionization cross section calculations, both for electron and proton impact
ionization.
Among the compilations subject to analysis, EADL (Evaluated Data Library)
\cite{eadl} contributes to deteriorate the accuracy of ionization cross sections
with respect to empirical compilations.

Characteristic K and L-shell X-ray transition energies are more accurately
calculated by using binding energies compiled by Larkins.
No significant effect depending on the choice of binding energies is observed in
the photon spectrum in Compton scattering accounting for Doppler broadening.

\section{Conclusions}

The exploratory analysis of epistemic uncertainties in two Monte Carlo
simulation domains has highlighted their contribution to simulation accuracy and
their capability of generating systematic effects in simulation results.
Further investigations are in progress to identify and quantify epistemic
uncertainties in Geant4 physics models.

Due to the limited page allocation in these conference proceedings, the detailed
results of the analysis cannot be reported here; they can be found in dedicated publications
\cite{tns_bragg, tns_binding}.
%

\end{document}